\documentclass[letterpaper,journal]{IEEEtran}
\usepackage{amsmath,amsfonts,amssymb}
\usepackage{xcolor}
\usepackage{bm}
\usepackage{algorithmic}
\usepackage{array}
\usepackage[caption=false,font=normalsize,labelfont=sf,textfont=sf]{subfig}
\usepackage{textcomp}
\usepackage{stfloats}
\usepackage{url}
\usepackage{verbatim}
\usepackage{graphicx}
\def\BibTeX{{\rm B\kern-.05em{\sc i\kern-.025em b}\kern-.08em
    T\kern-.1667em\lower.7ex\hbox{E}\kern-.125emX}}
\usepackage{balance}
\usepackage{cite}
\PassOptionsToPackage{ruled, linesnumbered}{algorithm2e}
\usepackage{algorithm2e}
\allowdisplaybreaks
\usepackage{amsthm, amssymb}
\newtheorem{proposition}{Proposition}
\newtheorem{lemma}{Lemma}
\newtheorem{theorem}{Theorem}
\newtheorem{remark}{Remark}

\newcommand{\herm}{^{\mathsf{H}}}
\newcommand{\trans}{^{\mathsf{T}}}

\DeclareMathOperator{\tr}{\mathsf{tr}}

\DeclareMathOperator{\maximize}{maximize}
\DeclareMathOperator{\st}{subject~to}

\renewcommand{\baselinestretch}{0.98}

\begin{document}
\title{\Large{Flexible Intelligent Metasurface for Downlink Communications under Statistical CSI}}
\author{Vaibhav~Kumar,~\IEEEmembership{Member,~IEEE,} Anastasios~Papazafeiropoulos,~\IEEEmembership{Senior~Member,~IEEE,} Pandelis~Kourtessis,  John~Senior, Marwa~Chafii,~\IEEEmembership{Senior~Member,~IEEE,} Dimitra~I.~Kaklamani, and Iakovos~S.~Venieris
\thanks{The work of Vaibhav Kumar and Marwa Chafii was supported by the Center for Cyber Security through New York University Abu Dhabi Research Institute under Award G1104. The work of Marwa Chafii was also supported by Tamkeen under the Research Institute NYUAD grant CG017.}
\thanks{Vaibhav~Kumar and Marwa~Chafii are with the Wireless Research Lab, Engineering Division, New York University Abu Dhabi, UAE. Marwa Chafii is also with NYU WIRELESS, NYU Tandon School of Engineering, New York, USA (e-mail: vaibhav.kumar@ieee.org; marwa.chafii@nyu.edu).} 
\thanks{Anastasios Papazafeiropoulos,  Pandelis Kourtessis, and John Senior are with the Communications and Intelligent Systems Research Group, University of Hertfordshire, Hatfield AL10 9AB, U.K. (e-mail: tapapazaf@gmail.com, p.kourtessis@herts.ac.uk).}
\thanks{Dimitra I. Kaklamani is with the Microwave and Fiber Optics Laboratory, and Iakovos S. Venieris is  with the Intelligent Communications and Broadband Networks Laboratory, School of Electrical and Computer Engineering, National Technical University of Athens, Zografou, 15780 Athens,	Greece.}}


\maketitle

\begin{abstract}
Flexible intelligent metasurface (FIM) is a recently developed, groundbreaking hardware technology with promising potential for 6G wireless systems. Unlike conventional rigid antenna array (RAA)-based transmitters, FIM-assisted transmitters can dynamically alter their physical surface through morphing, offering new degrees of freedom to enhance system performance. In this letter, we depart from prior works that rely on instantaneous channel state information (CSI) and instead address the problem of average sum spectral efficiency maximization under \emph{statistical} CSI in a FIM-assisted downlink multiuser multiple-input single-output setting. To this end, we first derive the spatial correlation matrix for the FIM-aided transmitter, and then propose an iterative FIM optimization algorithm based on the gradient projection method. Simulation results show that with statistical CSI, the FIM-aided system provides a significant performance gain over its RAA-based counterpart in scenarios with strong spatial channel correlation, whereas the gain diminishes when the channels are weakly correlated. 
\end{abstract}


\section{Introduction}
\IEEEPARstart{T}{he} rise of industrial 6G applications such as integrated sensing and communication, digital twins, extended reality, and autonomous systems requires very reliable and low latency links. Although high frequency bands (for example, 26, 28, and 39\,GHz) offer large bandwidth~\cite{Ahmed2025UpperMidBand}, their coverage and deployment challenges mean that sub-6\,GHz bands will remain important, and new physical layer mechanisms are needed to further improve coverage and capacity.

Flexible intelligent metasurfaces (FIMs) have recently emerged as a promising such mechanism. Built from flexible metamaterials~\cite{22_Nature_FIM}, FIMs can be passively or actively morphed in three dimensions~\cite{25_TAP_Jiancheng}, so that both the radiation pattern and the array geometry become design variables. This additional degree of freedom has been shown to reduce transmit power in multiuser MISO downlink~\cite{25_TWC_Jiancheng}, improve achievable rate in point to point MIMO~\cite{25_TCOM_Jiancheng_MIMO}, and enhance multi target sensing performance by controlling spatial correlation between steering vectors at different angles~\cite{25_TVT_Jiancheng_Sensing}. These results indicate that FIM based transmit architectures are a strong candidate for future 6G systems.

It is important to note that all existing works on FIM-aided systems assume \emph{instantaneous} channel state information (CSI) for performance optimization and a center frequency of $28$~GHz~\cite{25_TWC_Jiancheng, 25_TCOM_Jiancheng_MIMO, 25_TVT_Jiancheng_Sensing}. The reported performance gains over RAA-assisted systems are primarily attributed to the ability of surface morphing to enhance channel gains and improve interference suppression. It has explicitly been noted in~\cite{25_TCOM_Jiancheng_MIMO} that further research is required on the accurate channel estimation in FIM-aided systems to achieve the performance benefits. Motivated by this, in this letter, we consider a FIM-aided downlink MISO system with spatially correlated channels in the sub-$6$~GHz band, where only \emph{statistical} CSI is available at the transmitter. The main contributions of this letter are listed as follows:
\begin{itemize}
	\item We consider a FIM-assisted downlink MU-MISO system under statistical CSI, where we derive the spatial correlation matrix corresponding to the morphable surface, propose an MMSE-based channel estimation method, and obtain a closed-form expression for the average sum spectral efficiency (SE).
	
	\item To maximize the average sum SE, we formulate a non-convex optimization problem and develop an efficient solution based on the projected gradient method (PGM), which iteratively updates the FIM configuration.
	
	\item Through simulations, we demonstrate the performance superiority of the FIM-assisted system over a conventional RAA baseline under correlated channels, highlighting the impact of key parameters including morphing range, spatial correlation, number of users, and number of transmit elements.
\end{itemize}

\section{System Model}
\begin{figure}[t]
	\begin{center}
		\includegraphics[width=0.5\linewidth]{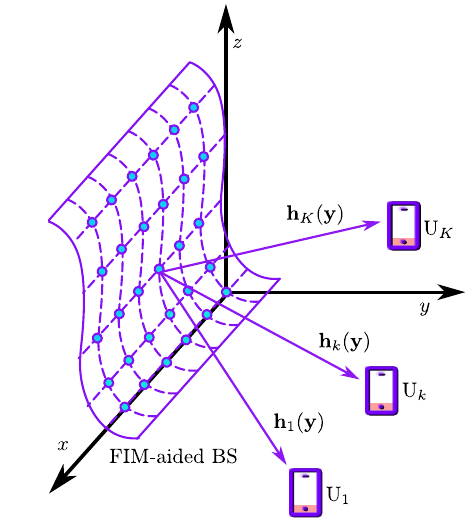}
		\caption{A large FIM-aided system with multiple users.}
		\label{fig:SystemModel}
	\end{center}
\end{figure}

We consider a multiuser MISO communication system, shown in Fig.~\ref{fig:SystemModel}, where a BS is deployed with a large FIM and serves $K$ single-antenna users, in an isotropic scattering environment. In particular, the antenna array, consisting of the FIM is a uniform planar array (UPA) located on the $x-z$ plane with $N=N_{x}N_{z}$ transmitting elements, where $N\ge K$ while $N_{x}$ and $N_{z}$ correspond to the number of antenna elements along the $x$- and $z$-axes, respectively.   For the sake of exposition, we denote $\mathcal{N}=\{1, 2,\ldots, N \}$ and $\mathcal{K}=\{1, 2,\ldots , K\}$. Note that the FIM is placed along the $x-z$ plane for notational simplicity; the results derived in this letter apply for any array orientation.

Contrary to rigid antenna arrays used in conventional communication systems, the radiating elements of the FIM can be flexibly placed perpendicularly to the surface, i.e., at the $y$-axis by means of a controller, as illustrated in the three-dimensional (3D) space in Fig.~\ref{fig:SystemModel}. Therein, $\theta$ is the elevation angle and $\varphi$ is the azimuth angle. Given that the communication takes place in an isotropic scattering environment, the multipath components in front of the FIM follow a uniform distribution that is expressed by $f(\theta, \varphi)=\cos(\theta) / 2\pi, \theta\in [-\frac{\pi}{2}, \frac{\pi}{2}],\varphi\in [-\frac{\pi}{2}, \frac{\pi}{2}]$. The implementation of the antenna elements takes place row-by-row, and  the location of the $n$-th element with respect to the first element as a reference point found at the origin, is given according to Fig.~\ref{fig:SystemModel} as $\mathbf u_{n}=[x_{n}, y_{n},  z_{n}]\trans\in \mathbb{R}^{3} , n \in\mathcal{N}$, 
with  $x_{n} = \mod(n-1, N_{x})d_{\mathrm{H}}$ and $z_{n} = \lfloor(n-1)/N_{x}\rfloor d_{\mathrm{V}}$ being the horizontal and vertical indices of element $n$, respectively. Also, $d_{\mathrm{H}}$ and $d_{\mathrm{V}}$ denote  the spacing between 	adjacent antenna elements in the $x$- and $z$-directions. Moreover, the $y$-coordinate of the $n$-th radiating element should satisfy $y_{\min}\le y_{n}\le y_{\max},  n \in\mathcal{N}$, 
where $y_{\max}$ and $y_{\min}$ are the maximum and minimum $y$-coordinates of each element. For the sake of convenience, we set $y_{\min}=0$~\cite{25_TWC_Jiancheng}. The surface shape of the FIM  is given by $\mathbf y=[y_{1}, \ldots, y_{N}]\trans \in \mathbb{R}^{N}$. Also, we define the morphing range $\zeta = y_{\max}-y_{\min}>0$, which characterizes the level of deformation.\footnote{We note that the adopted displacement model only increases inter-element distances, ensuring that array morphing does not exacerbate mutual coupling or introduce sparsity artifacts.}

In this case, the array response vector of a plane wave is then given by $\mathbf a (\mathbf y,\theta, \varphi)=\big[\exp\big\{{j  \mathbf k  (\theta, \varphi)\trans \mathbf u_{1}}\big\}, \ldots, \exp\big\{{j \mathbf k  (\theta, \varphi)\trans \mathbf u_{N}}\big\}\big]\trans$,
where $\lambda$ is the wavelength of the transmit wave  having  elevation angle $\theta$ and azimuth angle $\varphi$, and $\mathbf k( \theta, \varphi)\in \mathbb{R}^{3}$ denotes the wave vector, given as~(c.f.~\cite{21_Emil_RayleighStatistical}) $\mathbf k( \theta, \varphi)=\frac{2 \pi}{\lambda}[\cos(\theta)\cos(\varphi),\cos(\theta)\sin(\varphi), \sin(\theta) ]\trans$. 

Let $L$ be the number of plane waves; then the channel between the FIM and the $k$-th user is expressed as $\mathbf h_{k} (\mathbf y) = \sum \nolimits_{l=1}^{L}\frac{c_{k,l}}{\sqrt{L}} \mathbf a (\mathbf y, \theta_{l}, \varphi_{l})$, 
where $c_{k,l}/\sqrt{L} \in \mathbb{C}$ is the complex signal attenuation of the $l$-th component toward $k$-th user. We assume that $c_{k,l}\ \forall k, l,$ are 
independent and identically distributed (i.i.d.) with zero mean and variance $A\mu_k$, where $A$ is the area of a FIM antenna element and  $\mu_{k}$ is the average intensity
attenuation. According to the central limit theorem, as $L \to \infty$, we obtain that $\mathbf h_{k}$ converges  in distribution as $\lim_{L \to \infty} \mathbf h_k \sim  \mathcal{CN}(\boldsymbol{0}, \mathbf R_{k} (\mathbf y))$, 
where $\mathbf R_{k}(\mathbf y) = A \mu_{k}\mathbf R_{\mathrm{FIM}}(\mathbf y)$  with $\mathbf R_{\mathrm{FIM}}(\mathbf y) \in \mathbb{C}^{N \times N}$ being the normalized spatial correlation  of the FIM obtained as
\begin{align}
	\mathbf R_{\mathrm{FIM}}(\mathbf y) = & \ \tfrac{\mathbb E \{\mathbf h_k (\mathbf y) \mathbf h_k \herm (\mathbf y)\}}{A \mu_k}  = \mathbb E \{\mathbf a (\mathbf y, \theta, \varphi) \mathbf a\herm (\mathbf y, \theta, \varphi)\}. 
\end{align}
Generally, the $(n,m)$-th element of $\mathbf R_{\mathrm{FIM}}(\mathbf y)$ can be written as
\begin{align}
	& \ [\mathbf R_{\mathrm{FIM}}(\mathbf y)]_{n,m} \notag \\
	= & \ \mathbb E\{\exp\{\mathbf k  (\theta, \varphi)\trans (\mathbf u_{n} - \mathbf u_{m})\}\}  = \mathbb E \{\exp(j\tfrac{2\pi}{\lambda}B)\}, \label{correlation1}
\end{align}
where $B=(x_{n}-x_{m})\cos(\theta)\cos(\varphi)+y_{n}\cos(\theta)\sin(\varphi)+(z_{n}-z_{m})\sin(\theta)$ and  $y_{m}=0$ since we want to express it as $\mathbf R_{\mathrm{FIM}}(\mathbf y)$, i.e.,  in terms of  the $y$-coordinates with respect to the origin. A closed-form expression for $\mathbf R_{\mathrm{FIM}}(\mathbf y)$ is obtained in the following proposition. 

\begin{proposition}\label{CorrelationProof}
	Under conditions of isotropic scattering, the spatial correlation matrix $\mathbf R_{\mathrm{FIM}}(\mathbf y)$ is modeled by means of its elements as
	\begin{align}
		[\mathbf R_{\mathrm{FIM}}(\mathbf y)]_{n,m} = \mathrm{sinc}\big( 2 \pi\|\mathbf u_{n} - \mathbf u_{m} \| / \lambda \big).\label{correlation}
	\end{align}
\end{proposition}
\begin{proof}
	See Appendix~\ref{CorrelationProof1}.	
\end{proof}
It is important to emphasize that, although~\eqref{correlation} resembles~\cite[eqn.~(11)]{21_Emil_RayleighStatistical}, the correlation matrix in the latter is constant (i.e., fixed for all realizations), whereas in the case of FIM it depends on the surface morphing vector ${\mathbf{y}}$.
\subsection{FIM Channel Estimation}\label{ChannelEstimation}
This section presents the channel estimation procedure performed at the BS to obtain the estimated channel vector $\widehat{\mathbf h}_{k}$ between the $k$-th user and the FIM. A narrowband quasi-static block fading model is considered, where each coherence block has duration $\tau_{\mathrm c}$ channel uses, and the uplink training occupies $\tau$ channel uses. To this end, we apply the minimum mean-square error (MMSE) approach.\footnote{This approach is widely used in conventional mMIMO systems~\cite{Marzetta2016}, but it is applied here for the first time to FIM systems. Although the channel estimation structure appears similar to that in mMIMO, it differs fundamentally because $\mathbf R_k(\mathbf y)$ depends on $\mathbf y$.}

During the training phase, all users simultaneously transmit mutually orthogonal pilot sequences. The pilot sequence of user $k$, denoted by $\mathbf x_k = [x_{k,1}, \ldots, x_{k,\tau}]^{\mathrm H} \in \mathbb{C}^{\tau \times 1}$, has length $\tau \ge K$ symbols and satisfies $\|\mathbf x_k\|^2 = \tau p_{\mathrm{train}}$, where $p_{\mathrm{train}}$ is the per-symbol transmit power. We also assume constant-modulus pilot symbols such that $|x_{k,i}|^2 = p_{\mathrm{train}}$ for all $k,i$. Moreover, orthogonality implies $\mathbf x_k^{\mathrm H} \mathbf x_i = 0$ for all $k \neq i$.

The superimposed received pilot matrix at the BS is $\mathbf Y_{\mathrm{train}}(\mathbf y) = \sum \nolimits_{i\in\mathcal K} \mathbf h_{i}(\mathbf y)\mathbf x_{i}^{\mathrm H}
+ \mathbf Z_{\mathrm{train}}$, 
where $\mathbf Z_{\mathrm{train}} \in \mathbb{C}^{N\times\tau}$ has independent columns distributed as $\mathcal{CN}(\mathbf 0, \sigma^2 \mathbf I)$. Multiplying $\mathbf Y_{\mathrm{train}}(\mathbf y) $ with $\mathbf x_k / (\tau p_{\mathrm{train}})$ yields $	\mathbf r_k(\mathbf y) 
= \frac{1}{\tau p_{\mathrm{train}}}\mathbf Y_{\mathrm{train}}(\mathbf y)\mathbf x_k = \mathbf h_k(\mathbf y) + \mathbf z_k$, 
where $\mathbf z_k \sim \mathcal{CN}\big(\mathbf 0, \frac{\sigma^2}{\tau p_{\mathrm{train}}}\mathbf I\big)$. Since $\mathbf h_k(\mathbf y)$ is Gaussian distributed, its MMSE estimate follows directly.

\begin{lemma}\label{PropositionDirectChannel}
	The MMSE estimated channel vector is given by
	\begin{align}
		\widehat{\mathbf h}_{k}(\mathbf y) = \mathbf R_{k}(\mathbf y) \mathbf Q_{k}(\mathbf y) \mathbf r_{k}(\mathbf y), \label{estim1}
	\end{align}
	where $\mathbf Q_{k}(\mathbf y) = \big(\mathbf R_{k}(\mathbf y) + \frac{\sigma^2}{ \tau p_{\mathrm{train}}} \mathbf I_{N}\big)^{-1}$. The estimation error is $\widetilde{\mathbf h}_{k}(\mathbf y) = \mathbf h_{k}(\mathbf y) - \widehat{\mathbf h}_{k}(\mathbf y)$ while it is independent of the estimate $\widehat{\mathbf h}_{k}$. Moreover, the MSE matrix is given by
	\begin{align}
		\mathbf{MSE}_{k}(\mathbf y) = \mathbf R_{k}(\mathbf y)-	\boldsymbol{\Psi}_{k}(\mathbf y), \label{mse}
	\end{align}
	where 
	\begin{align}
		\boldsymbol{\Psi}_{k}(\mathbf y) = \mathbb E \big\{\widehat{\mathbf h}_{k}(\mathbf y)	\widehat{\mathbf h}_{k}\herm (\mathbf y) \big\} = \mathbf R_{k}(\mathbf y) \mathbf Q_{k}(\mathbf y) \mathbf R_{k}(\mathbf y) .\label{var1}
	\end{align} 
\end{lemma}
\begin{proof}
	See Appendix~\ref{lem1}.	
\end{proof}
\begin{remark}
	In the FIM-assisted system, the transmit antenna positions are parameterized by the morphing vector $\mathbf y$, so that the spatial correlation matrix $\mathbf R_k(\mathbf y)$, and hence both the estimated-channel covariance $\boldsymbol{\Psi}_k(\mathbf y)$ and the MMSE error-covariance $\mathbf{MSE}_k(\mathbf y)$, become configuration-dependent. This direct coupling between array geometry and MMSE estimation statistics is a key difference from conventional RAA-assisted systems.
\end{remark}
\subsection{Downlink Transmission and Achievable Spectral Efficiency}
Given the CSI using the previous technique, and after application of linear precoding, the signal received at the $k$-th user is given by 
\begin{equation}
	r_{k}(\mathbf y) = {\mathbf h}\herm_{k}(\mathbf y) \mathbf x (\mathbf y) +  z_{k}, \label{DLreceivedSignal2}
\end{equation}
where $z_{k} \sim \mathcal{CN}(0,1)$ denotes the AWGN at the $k$-th user, and $\mathbf x(\mathbf y) = \sqrt{\eta(\mathbf y) P/K} \sum_{i \in \mathcal K} \mathbf f_{i}(\mathbf y)l_{i}$ denotes the transmit signal vector by the BS with a total power $P$. Moreover, $\mathbf f_{i}(\mathbf y) \in \mathbb C^{N \times 1}$ denotes the linear precoding vector for the $i$-th user, while $l_{i}$ is the  data symbol with $\mathbb E\{|l_{i}|^{2}\}=1$. Note that we have assumed \emph{equal power allocation} as commonly assumed in the massive MIMO literature~\cite{Hoydis2013}. Also, $\eta(\mathbf y)$  is a normalization parameters given by $\eta(\mathbf y) = \frac{K}{\mathbb E\{\tr (\mathbf F(\mathbf y) \mathbf F(\mathbf y)\herm) \}} $,
with $ \mathbf F(\mathbf y) = [\mathbf f_{1}(\mathbf y), \ldots, \mathbf f_{K}(\mathbf y)] \in\mathbb{C}^{N \times K}$. 

Next, following the use-and-then-forget (UaTF) bound~\cite{Bjoernson2017}, the downlink achievable average sum SE in nats/s/Hz is given by
\begin{equation}
	\mathsf{SE}(\mathbf y)	= \tfrac{\tau_{\mathrm{c}} - \tau}{\tau_{\mathrm{c}}} \sum \nolimits_{k \in \mathcal K}\ln \left ( 1+\gamma_{k}(\mathbf y)\right), \label{eq:rateExpression}
\end{equation}
with $\gamma_k(\mathbf y) = S_k(\mathbf y) / I_k (\mathbf y)$ represents the signal-to-interference-plus-noise ratio (SINR). Closed-form expressions for $S_k(\mathbf y)$ and $I_k(\mathbf y)$ can be given by 
\begin{align}
	S_k(\mathbf y) = |\mathbb E \{\mathbf h_k\herm(\mathbf y) \mathbf f_k (\mathbf y)\}|^2 = \tr^2(\boldsymbol{\Psi}_k(\mathbf y)), \label{eq:sigma_power_closed}
\end{align}
\begin{align}
	\!\!\!\! & I_k(\mathbf y) = \mathbb E \{ | \mathbf h_k \herm (\mathbf y) \widehat{\mathbf h}_k(\mathbf y) - \mathbb E \{\mathbf h_k \herm (\mathbf y) \widehat{\mathbf h}_k(\mathbf y)\}|^2 \}  \notag \\
	\!\!\!\! & \qquad \qquad \quad + \sum \nolimits_{i \in \mathcal K \setminus \{k\}} |\mathbb E\{\mathbf h_k\herm (\mathbf y) \mathbf f_i (\mathbf y)\}|^2 + \tfrac{K \sigma^2}{P \eta(\mathbf y)} \notag \\
	\!\!\!\! = \  & \tr \{\mathbf R_k (\mathbf y) \boldsymbol{\Psi}_{\mathrm{sum}} (\mathbf y)\} \! - \! \tr \{\boldsymbol{\Psi}^2_k(\mathbf y)\} \!+\! \tfrac{\sigma^2}{P} \tr\{\boldsymbol{\Psi}_{\mathrm{sum}}(\mathbf y)\}, 
\end{align}
where $\boldsymbol{\Psi}_{\mathrm{sum}}(\mathbf y) \triangleq \sum_{k \in \mathcal K} \boldsymbol{\Psi}_k (\mathbf y)$. We have omitted the detailed derivations of these closed-form expressions due to space constraints. 

\section{FIM Optimization}
With the given background, the problem of maximizing the average sum SE for the FIM-assisted downlink MISO system under statistical CSI can be formulated as follows: 
\begin{subequations}
	\label{eq:MainProblem}
	\begin{align}
		\underset{y_n \forall n \in \mathcal N}{\maximize}\  & \mathsf{SE}(\mathbf y), \label{eq:obj}\\
		\st\  & 0 \leq y_n \leq y_{\max} \ \forall n \in \mathcal N.  \label{eq:morphingConstraint}
	\end{align}
\end{subequations}
One can note that~\eqref{eq:MainProblem} is non-convex due to the non-convex objective. Since we are interested in the performance of large FIM systems in this letter, obtaining a low-complexity solution is crucial. For this purpose, we adopt a PGM-based approach. In particular, since~\eqref{eq:MainProblem} is always feasible, we start with a random initial morphing, and then ascent along the gradient direction, i.e., $\nabla_{y_n} \mathsf{SE}(\mathbf y)$ with a step-size obtained using Armijo's rule. The gradient $\nabla_{y_n} \mathsf{SE}(\mathbf y)$ is given in the following theorem. 
\begin{theorem} \label{grad_theorem}
	A closed-form expression for the gradient $\nabla_{y_n} \mathsf{SE}(\mathbf y)$ is given by 
	\begin{equation}
		\!\!\! \nabla_{y_n} \mathsf{SE}(\mathbf y) = \tfrac{\tau_{\mathrm c} - \tau}{\tau_{\mathrm c}} \sum \nolimits_{k \in \mathcal K} \tfrac{I_k (\mathbf y) \nabla_{y_n} S_k(\mathbf y) - S_k(\mathbf y) \nabla_{y_n} I_k (\mathbf y) }{(1 + \gamma_k(\mathbf y)) I_k (\mathbf y)^2},\!\! \label{eq:T1-1}
	\end{equation}
	where $\nabla_{y_n} S_k(\mathbf y) = 2 A \mu_k \tr (\boldsymbol{\Psi}_k) \tr (\mathbf{C}_k \dot{\mathbf{R}}_{n})$, and $\nabla_{y_n} I_k(\mathbf y) = A \mu_k \tr \big\{ \boldsymbol{\Psi}_{\mathrm{sum}} \dot{\mathbf R}_n \big\} + A \sum \nolimits_{i \in \mathcal K} \mu_i \tr \big\{ \mathbf D_{k,i} \dot{\mathbf R}_n \big\}  
	- 2 A \mu_k \tr \big(\mathbf E_k \dot{\mathbf R}_n \big) + \tfrac{\sigma^2}{P} A \sum \nolimits_{i \in \mathcal K} \mu_i \tr \big( \mathbf C_i \dot{\mathbf R}_n\big)$.
\end{theorem}
\begin{proof}
	See Appendix~\ref{grad_derivations}. 
\end{proof}

The PGM-based routine to obtain a stationary solution to~\eqref{eq:MainProblem} is outlined in~\textbf{Algorithm~\ref{algoPGM}}, where $\mathbf y^{(0)}$ is the initial random morphing, $\varkappa$ is the step-size (chosen via \emph{backtracking line search}), and $\Pi_{\mathcal Y_n}(x) \triangleq \max \{ \min\{x, \zeta\}, 0\}$ is the projection operator onto the feasible set of $y_n$, according to the maximum morphing range $\zeta$. 

\begin{algorithm}[t]
	\caption{The PGM algorithm to solve~\eqref{eq:MainProblem}}
	\label{algoPGM}
	\KwIn{ $\mathbf y^{(0)}, \varkappa$ }
	\KwOut{ $\mathbf y_{\mathrm{opt}}$}
	$\jmath \leftarrow1$\;
	\Repeat{convergence }{
		$y_n^{(\jmath)} = \Pi_{\mathcal Y_n}\big\{y_n^{(\jmath - 1)} + \varkappa \nabla_{y_n} \mathsf{SE}(\mathbf y^{(\jmath - 1)})\big\} \ \forall n \in \mathcal N$\;
		
		$n\leftarrow n+1$\;
	}
	$\mathbf y_{\mathrm{opt}} \leftarrow \mathbf y^{(\jmath)}$\;
\end{algorithm}

	\subsection{Convergence Analysis}
	\textbf{Algorithm~\ref{algoPGM}} employs a projected-gradient-based approach on the antenna morphing vector \(\mathbf y\), namely $\mathbf y^{(\jmath)} = \Pi_{\mathcal Y}\big\{ \mathbf y^{(\jmath - 1)} + \varkappa \nabla_{\mathbf y} \mathrm{SE}\big(\mathbf y^{(\jmath - 1)}\big)\big\}$, where \(\mathcal Y\) is the feasible morphing domain and $\varkappa$ is a suitably chosen step size. Since \(\mathrm{SE}(\mathbf y)\) is continuously differentiable and its gradient is Lipschitz continuous with constant \(L>0\), one can establish that every limit point of the sequence \(\{\mathbf y^{(\jmath)}\}\) is a stationary point of the optimization problem. In practice, we adopt a backtracking line search to select \(\varkappa\) so as to guarantee sufficient ascent in each iteration. Accordingly, the algorithm yields a non-decreasing sequence of objective values and converges in the sense that $\lim_{\jmath\to\infty}\|\mathbf y^{(\jmath+1)}-\mathbf y^{(\jmath)}\| = 0$,
	while any cluster point satisfies the first-order optimality condition $\big\langle \nabla_{\mathbf y}\mathsf{SE}(\mathbf y_{\mathrm{opt}}), \bar{\mathbf y} - \mathbf y_{\mathrm{opt}} \big\rangle \le 0,\quad \forall\,\bar{\mathbf y}\in\mathcal Y$. Although the non-convex nature of the underlying problem precludes a global optimality guarantee, our numerical results consistently indicate convergence in a small number of iterations.

\subsection{Computational Complexity}
In each iteration, constructing all covariance matrices $\mathbf R_k(\mathbf y)$ and the associated MMSE-related matrices $\boldsymbol\Psi_k (\mathbf y)$ requires $\mathcal{O}(K N^{3})$ operations, while the gradient evaluation exploiting the row-sparse structure of $\dot{\mathbf R}_n(\mathbf y)$ incurs $\mathcal{O}(K^{2} N^{2})$. Hence, the per-iteration complexity of the proposed algorithm is $\mathcal{O}(K N^{3} + K^{2} N^{2})$, which is dominated by $\mathcal{O}(K N^{3})$ in the practically relevant regime $N \gg K$. Since the proposed FIM-based morphing optimization relies solely on statistical CSI, it represents a long-term geometric design that does not need to be recomputed every channel coherence interval. Consequently, the one-time optimization cost is amortized over a long duration, making even a relatively high computational complexity acceptable in practice.

\section{Results and Discussion}
\begin{figure*}
	\centering
	\begin{minipage}{.32\textwidth}
		\centering
		\includegraphics[width=1\linewidth, height=4.8cm]{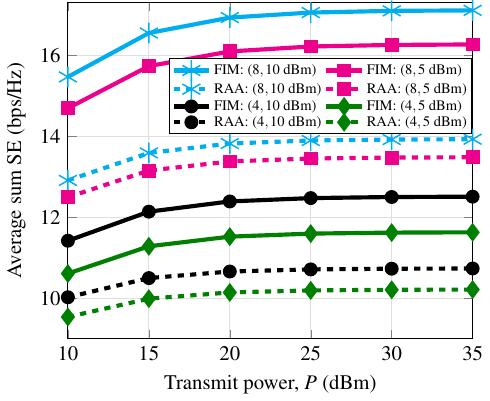}
		\caption{Impact of the transmit power on the average sum SE.}
		\label{fig:SE_vs_P}
	\end{minipage}%
	\hfill 
	\begin{minipage}{.32\textwidth}
		\centering
		\includegraphics[width=1\linewidth, height=4.8cm]{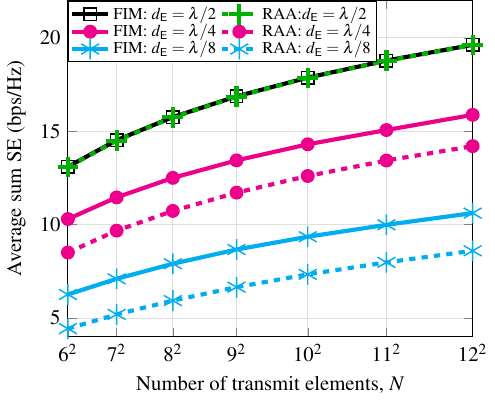}
		\caption{Impact of the number of transmit elements on the average sum SE.}
		\label{fig:SE_vs_N}
	\end{minipage}%
	\hfill 
	\begin{minipage}{.32\textwidth}
		\centering
		\includegraphics[width=1\linewidth, height=4.8cm]{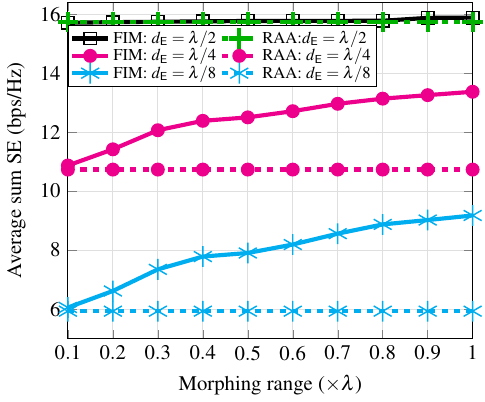}
		\caption{Impact of the morphing range on the average sum SE.}
		\label{fig:SE_vs_morphingRange}
	\end{minipage}
\end{figure*} 
We consider an FIM-assisted system where the downlink users are randomly distributed within a circle of radius $5$~m, centered at a distance of $100$~m from the FIM transmitter. The system is considered to be operating at a center frequency of $3.5$~GHz with $20$~MHz bandwidth. The noise power spectral density is $-174$~dBm/Hz and $d_{\mathrm H} = d_{\mathrm V} \triangleq d_{\mathsf E}$. Moreover, we assume $\tau_{\mathrm c} = 200$, $\tau = K$, and $p_{\mathrm{train}} = 10$~dBm. The average sum SE is obtained by averaging the results over 100 randomly generated user locations. To assess the performance benefits of the FIM-aided system, we consider the RAA-aided system as a benchmark.\footnote{Although a comparison between the FIM-assisted system under statistical CSI and its counterpart under instantaneous CSI is of interest, conducting such an analysis lies beyond the present scope. We therefore regard this as an important direction for future investigation.}

In Fig.~\ref{fig:SE_vs_P}, we show the impact of the transmit power, $P$, on the average sum SE of the FIM-assisted system, for $N = 64$, $d_{\mathsf E} = \lambda/4$, and a morphing range of $\lambda/2$. In the figure legend, the numbers in the parentheses denote $(K, p_{\mathrm{train}})$. It is evident from the figure that the performance of both the FIM- and RAA-assisted systems improves with an increase in the transmit power due to an improvement in the SINR. However, the average sum SE saturates at large $P$ values, due to the equal power allocation for all the users leading to dominating inter-user interference. At the same time, the system performance also increases with an increase in the number of users due to higher multi-user diversity. Interestingly, the performance difference between the FIM- and RAA-assisted systems increases with increasing $K$. For example, at $P = 30$~dBm and $p_{\mathrm{train}} = 10$~dBm, the average sum SE for the FIM-aided system improves by $16.5$\% for $K = 4$, and by $22.8$\% for $K = 8$ compared to the RAA-aided system. It is important to note that the performance advantage of the FIM-aided system over its RAA-aided counterpart stems from the ability of optimal morphing in the FIM to enhance the spatial correlation matrix $\mathbf R_k(\mathbf y)$, which remains fixed in the RAA-aided system. At the same time, the impact of MMSE channel estimation is also clearly evident from the figure, as when $p_{\mathrm{train}}$ increases from $5$~dBm to $10$~dBm, the average achievable SE increases significantly for both the FIM-based and RAA-based systems. However, the relative gain of the FIM-aided scheme becomes more pronounced at higher $p_{\mathrm{train}}$, since the improved MMSE channel estimates enable more effective geometry-dependent covariance shaping, whereas the RAA's fixed array cannot exploit this additional training SNR.

Next, we show the impact of the number of transmit elements, $N$, on the average sum SE of the FIM- and RAA-aided systems in Fig.~\ref{fig:SE_vs_N}, for $P = 30$~dBm, $K = 4$, and a morphing range of $\lambda/4$. As expected, the system performance improves with an increase in $N$, for both systems, due to higher spatial diversity and beamforming gain. However, it can be noted from the figure that as the size of each of the transmit elements or the inter-element size increases, the performance difference between the FIM- and RAA-based systems decreases. For example, at $N = 8^2$~dBm, the average sum SE for the FIM-aided system improves by $32.3$\% for $d_{\mathrm E} = \lambda/8$, by $16.5$\% for $d_{\mathrm E} = \lambda/4$, and by only $0.11$\% for $d_{\mathrm E} = \lambda/2$ compared to the RAA-aided system. The \emph{negligible} gain at $d_{\mathrm E} = \lambda/2$ arises because, for such an element spacing, the diagonal entries of $\mathbf{R}_{\mathrm{FIM}}(\mathbf{y})$ become much larger than the off-diagonal ones, making $\mathbf{R}_{\mathrm{FIM}}(\mathbf{y})$ behave almost like an identity matrix. Consequently, the spatial channel becomes effectively uncorrelated, leaving little room for optimization. Another interesting observation from the figure is that, for a fixed total transmitter size, one can either deploy $6^2$ elements with $d_{\mathsf E} = \lambda/2$ or nearly $12^2$ elements with $d_{\mathsf E} = \lambda/4$. For the FIM-aided system, the latter yields a $21.2$\% performance gain, indicating that it is preferable to pack smaller elements when the transmitter size is fixed. This is because having more, smaller elements provides finer spatial resolution and richer correlation diversity, allowing the FIM to exploit spatial channel variations more effectively. 

Finally, we examine the influence of the morphing range on the average sum SE of the FIM-assisted system in Fig.~\ref{fig:SE_vs_morphingRange}, for $P = 30$~dBm, $K = 4$, and $N = 64$. As the RAA-assisted system employs a fixed array geometry, its performance remains constant. In contrast, the SE of the FIM-aided system improves with an increased morphing range, attributed to the enhanced spatial degrees of freedom and the enlarged feasible set for optimizing the user-specific spatial correlation matrices $\mathbf{R}_k(\mathbf{y})$. This expanded design flexibility enables more effective spatial separation of user channels and improved interference management. However, practical morphing ranges are constrained by hardware, with existing flexible or mechanically reconfigurable metasurface prototypes typically supporting sub-wavelength to fractional-wavelength displacements. A detailed study of hardware-driven morphing limits is beyond the scope of this work but constitutes an important direction for future research.

\section{Conclusion}
This letter studied a FIM-assisted downlink MU-MISO system under statistical CSI in spatially correlated channels. We derived the spatial correlation matrix for the morphable FIM architecture, proposed an MMSE channel estimation scheme, and obtained a closed-form expression for the average sum SE. A PGM-based algorithm was developed to iteratively optimize the FIM shape. Simulation results confirmed the consistent performance gains of the FIM-assisted system over its RAA-based counterpart, with improvements increasing alongside the number of users, morphing range, and degree of correlation. These findings underscore the promise of FIM technology in enabling additional spatial design flexibility for future wireless systems.

\appendices
\section{Proof of Proposition~\ref{CorrelationProof}}\label{CorrelationProof1}	
For the proof, we start by considering two FIM elements, denoted as $n$ and $m$, placed on the same row obeying $x_{n}=x_{m}$ and $y_{n}=y_{m}$, which gives $(z_{n}-z_{m})d_{\mathrm{V}}=\|\mathbf u_{n}-\mathbf u_{m}\|$. To this end,  $[\mathbf R_{\mathrm{FIM}}(\mathbf y)]_{n,m}$ in \eqref{correlation1} is derived as
\begin{align}
	& [\mathbf R_{\mathrm{FIM}} (\mathbf y)]_{n,m} = \!  \int_{-\pi/2}^{\pi/2} \! \int_{-\pi/2}^{\pi/2} \!\! e^{j \frac{2 \pi }{\lambda}\|\mathbf u_{n} - \mathbf u_{m} \| \sin(\theta)}f(\theta, \varphi) \operatorname{d} \theta \operatorname{d}\varphi \nonumber \\
	&  =  \int_{-\pi/2}^{\pi/2} 0.5 e^{j \frac{2 \pi }{\lambda} \| \mathbf u_{n} - \mathbf u_{m}\| \sin(\theta)} \cos(\theta)\operatorname{d}\theta \notag \\
	& = \tfrac{\sin\left(2 \pi\|\mathbf u_{n}-\mathbf u_{m}\| / \lambda\right)}{\left(2 \pi\|\mathbf u_{n}-\mathbf u_{m}\| / \lambda\right)} = \mathrm{sinc}\left(2 \pi\|\mathbf u_{n}-\mathbf u_{m}\| / \lambda\right). \label{proof2}
\end{align}
This concludes the proof.

\section{Proof of Lemma~\ref{PropositionDirectChannel}}\label{lem1}
We can directly apply the results in \cite{Kay} to derive the MMSE channel estimate of  ${\mathbf h}_{k}$. Thus, the MMSE estimate of $\mathbf h_{k}$ is given due to~\cite[Eq. 15.64]{Kay} as
\begin{align}
	\widehat{\mathbf h}_{k} (\mathbf y) = \mathbb E \big\{\mathbf r_{k}(\mathbf y) \mathbf h_{k}\herm (\mathbf y) \big\} \left(\mathbb E\!\left\{\mathbf r_{k}(\mathbf y) \mathbf r_{k}\herm (\mathbf y) \right\} \right) ^{-1} \mathbf r_{k}(\mathbf y),\label{Cor6}
\end{align}
where $\mathbb E\left\{\mathbf r_{k}(\mathbf y) \mathbf h_{k}\herm (\mathbf y)\right\} = \mathbb E\left\{\mathbf h_{k} (\mathbf y) \mathbf h_{k}\herm (\mathbf y)\right\} = \mathbf R_{k} (\mathbf y)$
because $\mathbf r_{k} (\mathbf y)$  and $\mathbf h_{k}(\mathbf y)$ are uncorrelated. Also, we have $\mathbf Q_{k}^{-1} (\mathbf y) \triangleq \mathbb E\left\{\mathbf r_{k} (\mathbf y) \mathbf r_{k}\herm (\mathbf y)\right\} = \mathbf R_{k} (\mathbf y) + (\sigma^2/ (\tau p_{\mathrm{train}})) \mathbf I_{N}$. 

The covariance matrix of the estimated channel is then obtained as $\boldsymbol{\Psi}_{k} (\mathbf y) \triangleq \mathbb E \big\{\widehat{\mathbf h}_{k}(\mathbf y)	\widehat{\mathbf h}_{k}\herm (\mathbf y) \big\} = \mathbf R_{k} (\mathbf y) \mathbf Q_{k} (\mathbf y) \mathbf R_{k}(\mathbf y)$.
The estimation error $ \widetilde{\mathbf h}_{k} $ and channel estimate $ \widehat{\mathbf h}_{k} $ are independent of each other due to the orthogonality principle. Also, the covariance matrix of  $\widetilde{\mathbf h}_{k}$ is $\mathbf{MSE}_{k}(\mathbf y) = \mathbf R_{k}(\mathbf y) - \mathbf R_{k} (\mathbf y) \mathbf Q_{k}(\mathbf y) \mathbf R_{k}(\mathbf y)$, 
which concludes the proof.

\section{Proof of Theorem~\ref{grad_theorem}}\label{grad_derivations}
We start with the derivation of the gradient of the achievable sum SE with respect to $y_{n}$. For the sake of convenience, we omit the dependence on $y_{n}$. The gradient of the sum SE w.r.t. $y_n$ can be given by $\nabla_{y_n} \mathsf{SE} = \tfrac{\tau_{\mathrm c} - \tau}{\tau_{\mathrm c}} \sum \nolimits_{k \in \mathcal K} \tfrac{I_k \nabla_{y_n}S_k - S_k \nabla_{y_n} I_k}{(1 + \gamma_k) I_k^2}$.
Next, one can obtain $\nabla_{y_n}S_k$ as $\nabla_{y_n} S_k = \nabla_{y_n} \tr^2 (\boldsymbol{\Psi}_k) = 2 \tr (\boldsymbol{\Psi}_k) \tr (\nabla_{y_n} \boldsymbol{\Psi})
=  2 \tr (\boldsymbol{\Psi}_k) \tr (\nabla_{y_n} \{\mathbf{R}_k \mathbf{Q}_k \mathbf{R}_k\})  
= 2 A \mu_k \tr (\boldsymbol{\Psi}_k) \tr (\mathbf{C}_k \dot{\mathbf{R}}_{n})$, 
where $\mathbf{C}_k \triangleq \mathbf Q_k \mathbf R_k - \mathbf Q_k \mathbf R_k^2 \mathbf Q_k + \mathbf R_k \mathbf Q_k$ and $\dot{\mathbf{R}}_{k, n}$ is an all-zero matrix, except the $n$-th row whose $m$-th element is given by 
\begin{equation}
	\!\!\!\! \dot{\mathbf R}_n = \left[ \tfrac{\cos(2\pi d_{nm} / \lambda )}{(2\pi d_{nm} / \lambda)} - \tfrac{\sin (2\pi d_{nm} / \lambda)}{(2\pi d_{nm}^2 / \lambda)} \right] \tfrac{(y_n - y_m)}{d_{nm}}, \label{eq:grad_proof-C}
\end{equation}
provided $n \neq m$ and $d_{n,m} = \|\mathbf u_n - \mathbf u_m\|$. 

Next, we have $\nabla_{y_n} I_k = \nabla_{y_n} \tr \! \big\{ \mathbf R_k \boldsymbol{\Psi}_{\mathrm{sum}} \big\} \!-\! \tr \! \big\{\boldsymbol{\Psi}^2_k \big\} \!+\! \frac{\sigma^2}{P} \tr \! \big\{ \boldsymbol{\Psi}_{\mathrm{sum}} \big\} 
=  \nabla_{y_n} \big( I_{k,1} - I_{k, 2} + I_{k, 3}\big)$. 
A closed-form expression for $\nabla_{y_n} I_{k, 1}$ can be obtained as $\nabla_{y_n} I_{k, 1} =  \nabla_{y_n} \tr \big\{ \mathbf R_k \boldsymbol{\Psi}_{\mathrm{sum}} \} 
=  \tr \big\{ \big(\nabla_{y_n} \mathbf R_k \big) \boldsymbol{\Psi}_{\mathrm{sum}} \big\} + \tr \big\{ \mathbf R_k \big(\nabla_{y_n} \boldsymbol{\Psi}_{\mathrm{sum}}\big)\big\} 
=  \tr \big\{ \boldsymbol{\Psi}_{\mathrm{sum}} \big(\nabla_{y_n} \mathbf R_k \big) \big\} + \sum \nolimits_{i \in \mathcal K} \tr \big[ \mathbf R_k \big\{ \big(\nabla_{y_n} \mathbf R_i \big) \mathbf Q_i \mathbf R_i  - \mathbf R_i \mathbf Q_i \big(\nabla_{y_n} \mathbf R_i \big) \mathbf Q_i \mathbf R_i + \mathbf R_i \mathbf Q_i \big(\nabla_{y_n} \mathbf R_i \big)\big\} \big] 
= \tr \big\{ \boldsymbol{\Psi}_{\mathrm{sum}} \big(\nabla_{y_n} \mathbf R_k \big) \big\} + \sum \nolimits_{i \in \mathcal K} \tr \big\{ \mathbf D_{k,i} \big(\nabla_{y_n} \mathbf R_i \big) \big\}
= A \mu_k \tr \big\{ \boldsymbol{\Psi}_{\mathrm{sum}} \dot{\mathbf R}_n \big\} + A \sum \nolimits_{i \in \mathcal K} \mu_i \tr \big\{ \mathbf D_{k,i} \dot{\mathbf R}_n \big\}$, 
where $\mathbf D_{k, i} \triangleq \mathbf Q_i \mathbf R_i \mathbf R_k - \mathbf Q_i \mathbf R_i \mathbf R_k \mathbf R_i \mathbf Q_i + \mathbf R_k \mathbf R_i \mathbf Q_i$. 
Similarly, we have $\nabla_{y_n} I_{k, 2} = \tr \big\{ 2 \boldsymbol{\Psi}_k \big(\nabla_{y_n} \boldsymbol{\Psi}_k \big) \big\} = 2 A \mu_k \tr \big(\mathbf E_k \dot{\mathbf R}_n \big)$, 
where $\mathbf E_k \triangleq \mathbf Q_k \mathbf R_k \boldsymbol{\Psi}_k - \mathbf Q_k \mathbf R_k \boldsymbol{\Psi}_k \mathbf R_k \mathbf Q_k + \boldsymbol{\Psi}_k \mathbf R_k \mathbf Q_k$. At the end, a closed-form expression for $\nabla_{y_n} I_{k, 3}$ can be given by $\nabla_{y_n} I_{k, 3} = \frac{\sigma^2}{P} A \sum \nolimits_{i \in \mathcal K} \mu_i \tr \big( \mathbf C_i \dot{\mathbf R}_n\big)$. 
This completes the proof. 
\bibliographystyle{IEEEtran}
\bibliography{ref}

@article{Ahmed2025UpperMidBand,
	title        = {Upper Mid‑Band Spectrum for {6G}: {V}ision, Opportunity and Challenges},
	author       = {Ahmad Bazzi and Roberto Bomfin and Marco Mezzavilla and Sundeep Rangan and Theodore Rappaport and Marwa Chafii},
	journal      = {arXiv preprint arXiv:2502.17914},
	year         = {2025},
	eprint       = {2502.17914},
	archivePrefix= {arXiv},
	primaryClass = {eess.SP},
	doi          = {10.48550/arXiv.2502.17914},
	note         = {version 2, revised May 12 2025}
}

@article{22_Nature_FIM,
	title={A dynamically reprogrammable surface with self-evolving shape morphing},
	author={Bai, Yun and others},
	journal={Nature},
	volume={609},
	number={7928},
	pages={701--708},
	year={2022},
	publisher={Nature Publishing Group UK London}
}

@ARTICLE{25_TAP_Jiancheng,
	author  = {An, Jiancheng and Debbah, Mérouane and Cui, Tie Jun and Chen, Zhi Ning and Yuen, Chau},
	journal = {IEEE Trans. Antennas Propag.},
	title   = {Emerging Technologies in Intelligent Metasurfaces: {S}haping the Future of Wireless Communications},
	note= {{DOI: 10.1109/TAP.2025.3571069}},
	year= {Early Access, 2025},
}

@article{25_TWC_Jiancheng,
	author={An, Jiancheng and Yuen, Chau and Renzo, Marco Di and Debbah, Mérouane and Poor, H. Vincent and Hanzo, Lajos},
	journal={IEEE Trans. Wireless Commun.}, 
	title={Flexible Intelligent Metasurfaces for Downlink Multiuser {MISO} Communications}, 
	year={2025},
	volume={24},
	number={4},
	pages={2940-2955},
	doi={10.1109/TWC.2025.3526843}}

@ARTICLE{25_TCOM_Jiancheng_MIMO,
	author={An, Jiancheng and Han, Zhu and Niyato, Dusit and Debbah, Mérouane and Yuen, Chau and Hanzo, Lajos},
	journal={IEEE Trans. Commun.}, 
	title={Flexible Intelligent Metasurfaces for Enhancing {MIMO} Communications}, 
	year={Early Access 2025},
	note={{DOI: 10.1109/TCOMM.2025.3550318}}}

@ARTICLE{25_TVT_Jiancheng_Sensing,
	author={Teng, Zihao and An, Jiancheng and Gan, Lu and Al-Dhahir, Naofal and Han, Zhu},
	journal={IEEE Trans. Veh. Technol.}, 
	title={Flexible Intelligent Metasurface for Enhancing Multi-Target Wireless Sensing}, 
	year={Early Access. 2025},
	note={{DOI: 10.1109/TVT.2025.3584865}}}

@Book{Marzetta2016,
	author    = {Marzetta, Thomas L. and Larsson, Erik G. and Yang, Hong and Ngo, Hien Quoc},
	title     = {Fundamentals of Massive {MIMO}},
	year      = {2016},
	publisher = {Cambridge University Press},
	doi       = {10.1017/CBO9781316799895},
	place     = {Cambridge},
}

@Book{Kay,
	Title                    = {Fundamentals of Statistical Signal Processing: Estimation Theory},
	Author                   = {S. M. Kay},
	Publisher                = {Upper Saddle River: Prentice Hall PTR},
	Year                     = {1993}
}

@Article{Hoydis2013,
	author    = {Hoydis, J. and ten Brink, S. and Debbah, M.},
	title     = {Massive {MIMO} in the {UL/DL} of Cellular Networks: {H}ow Many Antennas Do We Need?},
	doi       = {10.1109/JSAC.2013.130205},
	issn      = {0733-8716},
	number    = {2},
	pages     = {160-171},
	volume    = {31},
	journal   = {IEEE J. Select. Areas Commun.},
	year      = {2013},
}

@Article{Bjoernson2017,
	author    = {Bj{\"o}rnson, Emil and Hoydis, Jakob and Sanguinetti, Luca},
	title     = {Massive {MIMO} networks: Spectral, energy, and hardware efficiency},
	number    = {3-4},
	pages     = {154--655},
	volume    = {11},
	journal   = {Foundations and Trends{\textregistered} in Signal Processing},
	publisher = {Now Publishers, Inc.},
	year      = {2017},
}

@ARTICLE{21_Emil_RayleighStatistical,
  author={Björnson, Emil and Sanguinetti, Luca},
  journal={IEEE Wireless Commun. Lett.}, 
  title={Rayleigh Fading Modeling and Channel Hardening for Reconfigurable Intelligent Surfaces}, 
  year={2021},
  volume={10},
  number={4},
  pages={830-834},
  doi={10.1109/LWC.2020.3046107}}

\end{document}